\documentclass[preprint,showpacs,preprintnumbers,amsmath,amssymb]{revtex4}

\usepackage{graphicx}
\usepackage{dcolumn}
\usepackage{bm}

\newcommand{\ea}{\end{eqnarray}}

\begin{document}
%



\title{An interacting quark-diquark model of baryons}

\author{E. Santopinto.}
\email{santopin@ge.infn.it}
\affiliation{INFN and Universit\`a di Genova,
via Dodecaneso 33, 16142 Genova, Italy}
\date{\today}

\begin{abstract}
A simple quark-diquark model of baryons with direct and exchange
interactions has been constructed. Spectrum and form factors have been calculated and compared with experimental data.  Advantages and disadvantages of the model are discussed.
\end{abstract}
\maketitle


The notion of diquark is as old as the quark model itself. Gell-Mann 
\cite{gell} mentioned the possibility of diquarks in his original paper on
quarks. Soon afterwards, Ida and Kobayashi \cite{ida} and Lichtenberg and
Tassie \cite{lich} introduced effective degrees of freedom of diquarks in
order to describe baryons as composed of a constituent diquark and quark.
Since its introduction, many articles have been written on this subject 
\cite{ans} up to the most recent ones \cite{Wilczek}.

Different phenomenological indications for diquark
correlations  have been
collected during the years, such as some regularities in hadron spectroscopy,
the $\Delta I=\frac{1}{2}$ rule in weak non-leptonic decays 
\cite{Neubert}, 
some regularities in parton distribution
functions \cite{Close} and in spin-dependent structure functions
\cite{Close}.
Finally, although the phenomenon of color superconductivity \cite{bailingwilczek} 
in quark dense matter can not be considered an argument in support of
diquarks in the vacuum, it is nevertheless of interest since it stresses the important role of Cooper pairs of color superconductivity, which are color
antitriplet, flavour antisymmetric, scalar diquarks.

The introduction of diquarks in hadronic physics has some
similarities to
that of correlated pairs in condensed matter physics (superconductivity
\cite{bcs}) and in nuclear physics (interacting boson model \cite{ibm}), where
 effective bosons emerge from pairs of electrons \cite{coop} and nucleons
\cite{ibm2} respectively. However, while the origin of correlated electron
and nucleon pairs is clear (the electron-phonon interaction in condensed
matter physics and the short-range pairing interaction in nuclear
physics),
the microscopic origin of diquarks is not completely clear and its
connection with the fundamental theory (QCD) not fully understood,
 apart from the cold asymptotic high dense baryon case, in which the quarks form a Fermi surface and perturbative gluon interactions support the existence of a diquark BCS state known as color superconductor \cite{bailingwilczek}.
We can only say that in perturbative QCD the color antitriplet, flavor
antisymmetric  scalar channel is favored by one gluon exchange
\cite{DeRujula} and in non-perturbative regime 
by instanton interactions \cite{Shuryak}. 
 Regarding the hadron spectrum we are interested 
in the role of diquark correlations in non perturbative QCD. In this respect 
it is interesting the discussion 
of non-perturbative short-range, spin and flavour
 dependent correlations in hadrons, given in Ref.\cite{Faccioli}
considering an Instanton Liquid Model and the comparison of some of these results with LQCD ones.  However, the question remains open and 
other lattice QCD calculations are needed in order to understand it fully.

Nonetheless, as in nuclear physics, one may attempt to correlate the data in
terms of a phenomenological model. In this article, we address this question
by formulating a quark-diquark model with explicit interactions, in
particular with a direct and an exchange interaction. We then study the
spectrum which emerges from this model and we start to calculate the
 form factors which have been measured, or will be measured at dedicated facilities (TJNAF,MAMI, ...).

We think of a diquark as two correlated quarks treated as a point-like object, thought this is a rough approximation of an extended effective boson degree of freedom.  


We assume that baryons are composed of a constituent quark, $q$, and a
constituent diquark, $q^{2}=Q^{2}$.
We consider only light baryons, composed of
(u, d, s) quarks, with internal group $SU_{s}(2)\otimes ~SU_{f}(3)\otimes
SU_{c}(3)$. Using the conventional notation of denoting spin by its value
and flavor and color by the dimension of the representation, the quark has
spin $s=\frac{1}{2}$, $F={\bf {3}}$ and $C={\bf {3}}$.
 The diquark must be ${\bf {\overline{3}}}$ of $SU_{c}(3)$ since the total
hadron must be colorless. This limits the possible $SU_{sf}(6)$
 representations for the diquark \cite{lich} to be only the 
 ${\bf {21}}$ of $SU_{sf}(6),$ which is symmetric and contains $%
s_{12}=0$, $F_{12}={\bf {\overline{3}}}$ and $s_{12}=1$, $F_{12}={\bf {6}}$ . 
 This is 
because we think of the diquark as two correlated quarks in an antisymmetric 
non-excited state.

As one can see from the simple multiplication of the Young diagrams 
associated with two fundamental representations of $SU_{sf}(6)$

\vspace{-0.5cm}
\begin{eqnarray}
\setlength{\unitlength}{1.0pt}
\begin{picture}(10,10)(0,0)
\thinlines
\put ( 0, 0) {\line (1,0){10}}
\put ( 0,10) {\line (1,0){10}}
\put ( 0, 0) {\line (0,1){10}}
\put (10, 0) {\line (0,1){10}}
\end{picture} \;\;\otimes\;\; 
\setlength{\unitlength}{1.0pt}
\begin{picture}(10,10)(0,0)
\thinlines
\put ( 0, 0) {\line (1,0){10}}
\put ( 0,10) {\line (1,0){10}}
\put ( 0, 0) {\line (0,1){10}}
\put (10, 0) {\line (0,1){10}}
\end{picture} &=&   
\setlength{\unitlength}{1.0pt}
\begin{picture}(20,20)(0,0)
\thinlines
\put ( 0, 0) {\line (1,0){20}}
\put ( 0,10) {\line (1,0){20}}
\put ( 0, 0) {\line (0,1){10}}
\put (10, 0) {\line (0,1){10}}
\put (20,0) {\line (0,1){10}}
\end{picture} \;\;\oplus\;\; 
\setlength{\unitlength}{1.0pt}
\begin{picture}(10,30)(0,5)
\thinlines
\put ( 0, 0) {\line (1,0){10}}
\put ( 0,10) {\line (1,0){10}}
\put ( 0,20) {\line (1,0){10}}
\put ( 0, 0) {\line (0,1){20}}
\put (10, 0) {\line (0,1){20}}
\end{picture} \\
\nonumber \\
{\bf {6}~\otimes ~{6}~} &=&{\bf {21}~~\oplus ~~{15}~,}
\end{eqnarray}
by keeping only the representation ${\bf {21}}$ of $SU_{sf}(6)$,
 as in the diquark case, one is
deleting in baryons those states obtained by combining the 
representation ${\bf {15}}$ with that of the remaining quark ${\bf 6}$, 
i.e. a  ${\bf 70}$ and  ${\bf 20}$.

If one treats only non-strange baryons, the quark is in the representation $%
{\bf {4}}$ of the Wigner $SU_{st}(4)$, with $s=\frac{1}{2}$ and $t=\frac{1}{2%
},$ and the diquark is the representations with spin $s_{12}=0$, isospin $%
~t_{12}=0,$ and spin $s_{12}=1$, isospin $t_{12}=1$, i.e. the symmetric
representation ${\bf {10}}$ of $SU_{st}(4)\supset ~SU_{s}(2)\otimes
SU_{t}(2) $. The situation for the internal degrees of freedom is summarized
in Table I.

\begin{table}[t]
\caption{Combinations of possible irreducible representations of $SU_{c}(3)$%
, $SU_{s}(2)$ and $SU_{f}(3)$ for the quark, $q$, and the diquark, $Q^{2}$.
The right-hand side of the Table also shows the combinations of the possible
 $SU_{s}(2)$ and $SU_{t}(2)$ representations if only non-strange
baryons are involved. }
\label{tab:rapp}
\vspace{-0.4cm}
\par
\begin{center}
\begin{tabular}{cccccccc}
\hline
\hline
  & $SU_{c}(3)$ & $SU_{sf}(6)$ & $SU_{s}(2)$ & $SU_{f}(3)$ &  $SU_{st}(4)$
& $SU_{s}(2)$ & $SU_{t}(2)$ \\ 
\hline
$q$ &   ${\bf {3}}$ & ${\bf {6}}$ & $s=\frac{1}{2}$ & ${\bf {3}}$ &   $%
{\bf {4}}$ & $s=\frac{1}{2}$ & $t=~\frac{1}{2}$ \\ 
$Q^2$ &   ${\bf {\overline{3}}}$ & ${\bf {21}}$ & $s_{12}=0$ & ${\bf {%
\overline{3}}}$ &   ${\bf {10}}$ & $s_{12}=0$ & $t_{12}=0$ \\ 
  & $~$ & $~$ & $s_{12}=1$ & ${\bf {6}}$ &  &   $s_{12}=1$ & $%
t_{12}=1$ \\ 
\hline
\hline
\end{tabular}
\end{center}
\end{table}

The two diquark configurations $s_{12}=0$, $F_{12}={\bf {\overline{3}}}$ and $%
s_{12}=1$, $F_{12}=\bf{6}$ are split, by (among other things) color magnetic
forces \cite{deswart}, as shown in Fig. 1. It means that we have two possible 
diquark configurations: a scalar diquark
($s_{12}=0$, $F_{12}={\bf {\overline{3}}}$, $C={\bf {\overline{3}}}$) and a vector
one ($s_{12}=1$, $F_{12}=\bf{6}$, $C={\bf {\overline{3}}}$). The scalar diquark 
(the ``good diquark'' in Wilczek and Jaffe's terminology) is favoured since it is at lower energy, and  will be the dominant configuration in the more stable states.

\begin{figure}[!ht]

{\includegraphics[width=7cm]{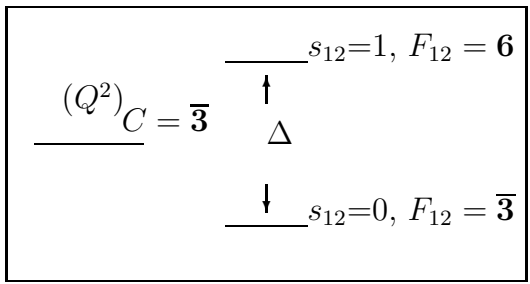}}

\vspace{-0.5cm}

\caption{\small Schematic picture of the split, between the two diquark configurations $s_{12}=0$, $F_{12}={\bf {\overline{3}}}$ and 
$s_{12}=1 $, $F_{12}={\bf{6}}$
.}
\label{flavor33e}
\end{figure}

We call the splitting between the two diquark configurations $\Delta $ and parametrize it as 
\begin{equation}
\Delta ~=~(B~+~C\delta_{0})~~,
\end{equation}
i.e. as a constant $B$ which acts equally in all states with $s_{12}=1$
plus a contact interaction which acts only on the ground state and of
strength C. The contact interaction emphasizes the different role that the
ground state has in a quark-diquark picture as compared with all other states.


Here, we consider a quark-diquark configuration in which the two constituents
are separated by a distance $r$. We use a potential picture and we introduce
a direct and an exchange quark-diquark interaction. For the direct term, we
consider a Coulomb-like plus a linear confining interaction 
\begin{equation}
V_{dir}(r)~=~-\frac{\tau }{r}~+~\beta r~~~.
\end{equation}
The importance of the Coulomb-like interaction was emphasized long ago by
Lipkin \cite{lip}. A simple mechanism that generates a Coulomb-like
interaction is one-gluon exchange. A natural candidate for the confinement
term is a linear one, as obtained in lattice QCD calculations and other
considerations \cite{cam}.

An exchange interaction is also needed, as emphasized by Lichtenberg \cite
{lichsec}. This is indeed the crucial ingredient of a quark-diquark
description of baryons.
We consider 
\begin{equation}
V_{ex}=(-1)^{l+1}2Ae^{-\alpha r}[\vec{s_{12}}\cdot \vec{s_{3}}+\vec{%
t_{12}}\cdot \vec{t_{3}}+2\vec{s_{12}}\cdot \vec{s_{3}} \vec{t_{12}}\cdot 
\vec{t_{3}}].
\end{equation}
Here $\vec{s_{12}}$, $\vec{s_{3}}$ and $\vec{t_{12}}$ and $\vec{t_{3}}$
 are the spin and the isospin operators of the diquark and the quark 
respectively.  
The sign $(-1)^{l}$ reflects the angular momentum dependence of the exchange
interaction.


A feature of the present approach, as compared with previous ones, is that
we attempt a simultaneous description of all states; indeed, in previous
 approaches the ground state has usually been excluded. It is certainly the case
that diquark correlations are particularly important for high $\ell $-states 
\cite{john} which have a more pronounced string-like behaviour. 
However, if one is interested in the overall
spectroscopy and also in the form factors (elastic and inelastic) that have been measured or will be measured (TJLAB, MAMI, ...) the ground state must be included in this
description.

Another important observation is that, in the quark-diquark model, $%
SU_{sf}(6)$ is badly broken and thus a classification of the spin-flavor
wave functions in terms of $SU_{sf}(6)$ is inappropriate. The appropriate
classification is that in terms of $SU_{s}(2)\otimes SU_{f}(3)$ 
\begin{equation}
|~s_{12},~F_{12};~\frac{1}{2},~{\bf 3};~S,~F\rangle ~~~,
\end{equation}
where $s_{12}$ and $F_{12}$ are, respectively, the spin and the flavor of the
diquark, $\frac{1}{2},{\bf 3}$ are the spin and flavor of the quark and $S$, 
$F$ are the total spin and flavor obtained by coupling the spin and flavor
of the diquark with those of the quark.

If only non-strange baryons are considered, one can use the isospin of the
diquark, $t_{12}$, and quark, $1/2$, and the total isospin, $T$. The
possible states are:

\begin{eqnarray}
|~s_{12}=~0,~t_{12}~=~0;~\frac{1}{2}~,~\frac{1}{2};~S~=~\frac{1}{2}, ~T~=~%
\frac{1}{2}~\rangle~~~,  \nonumber \\
|~s_{12}=~1,~t_{12}~=~1;~\frac{1}{2},~\frac{1}{2};~S~=~\frac{1}{2}, ~T~=~%
\frac{1}{2}~\rangle~~~,  \nonumber \\
|~s_{12}=~1,~t_{12}~=~1;~\frac{1}{2},~\frac{1}{2};~S~=~\frac{1}{2}, ~T~=~%
\frac{3}{2}~\rangle~~~,  \nonumber \\
|~s_{12}=~1,~t_{12}~=~1;~\frac{1}{2},~\frac{1}{2};~S~=~\frac{3}{2}, ~T~=~%
\frac{1}{2}~\rangle~~~,  \nonumber \\
|~s_{12}=~1,~t_{12}~=~1;~\frac{1}{2},~\frac{1}{2};~S~=~\frac{3}{2}, ~T~=~%
\frac{3}{2}~\rangle~~~.
\end{eqnarray}

The total wave functions are combinations of the spin-flavor wave functions
with the radial and orbital wave functions. These are 
very simple and straigthforward since the main advantage of the quark-diquark model is to reduce the baryon
problem (a three-body problem) to a two-body problem. The spatial part of
the wave functions is, for the central interaction of Eq. (4), 
\begin{equation}
\Psi_{n,l,m} (\vec{r})~=~R_{n,l}(r)~Y_{l,m}(\theta ,\varphi )~,
\end{equation}
where the radial wave function $R_{n,l}(r)$ can be obtained by solving the
radial equation. Although the numerical solution
poses no problem, we prefer to exploit here the special nature of the
interaction. For a purely Coulomb-like interaction the problem is
analytically solvable. The solution is trivial, with eigenvalues 
\begin{equation}
E_{n,l}~=~-\frac{\tau ^{2}m}{2~n^{2}}~~~~,~n~=~1,~2~...~~~~~.
\end{equation}
Here $m$ is the reduced mass of the diquark-quark configuration and $n$ the
principal quantum number. The eigenfunctions are the usual Coulomb functions 
\begin{equation}
R_{n,l}(r)=\sqrt{\frac{(n-l-1)!(2g)^{3}}{2n[(n+l)!]^{3}}}(2gr)^{l}~e^{-gr}L_{n-l-1}^{2l+1}(2gr),
\end{equation}
where for the associated Laguerre polynomials we have used the notation of Ref.
\cite{morse} and $g~=~\frac{\tau ~m}{n}$.

We treat all other interactions as perturbations. We begin with the linear
term. The matrix elements of $\beta ~r$ can be evaluated analytically, with
the result 
\begin{equation}
\Delta E_{n,l}=\int_{0}^{\infty }\beta r[R_{n,l}(r)]^{2}r^2dr=\frac{%
\beta }{2m\tau }[3n^{2}-l(l+1)].
\end{equation}
This perturbative estimate is only valid for small $n$ and $l$. (For large $%
n $ and $l$, the radial equation must be solved numerically.) Combining (9)
with (11), we can write the energy eigenvalues as 
\begin{equation}
E_{n,l}=-\frac{\tau ^{2}m}{2n^{2}}+\frac{\beta }{2m\tau }[3n^{2}-l(l+1)].
\end{equation}
\noindent Next comes the exchange interaction of Eq. (5). The spin-isospin part
is obviously diagonal in the basis of Eq. (7)
\begin{eqnarray}
\langle \vec{s}_{12}\cdot \vec{s}_{3}\rangle =\frac{1}{2}\left[
S(S+1)-s_{12}(s_{12}+1)-s_{3}(s_{3}+1)\right]  \nonumber \\
\langle \vec{t}_{12}\cdot \vec{t}_{3}\rangle =\frac{1}{2}\left[
T(T+1)-t_{12}(t_{12}+1)-t_{3}(t_{3}+1)\right] .
\end{eqnarray}
To complete the evaluation, we need the matrix elements of the exponential.
These can be obtained in analytic form 
\begin{equation}
I_{n,l}(\alpha )~=~\int_{0}^{\infty }~e^{-\alpha ~r}~[R_{n,l}(r)]^{2}r^2dr~~.
\end{equation}
The results are straightforward. Here, by way of example, we quote the result
for $l =n-1$ 
\begin{equation}
I_{n,l=n-1}(\alpha )~=~(\frac{1}{1+\frac{n~\alpha }{2\tau ~m}})^{2n+1}~~~.
\end{equation}


\noindent Combining all pieces together, the Hamiltonian is 
\begin{eqnarray}
H =E_{0}+\frac{p^{2}}{2m}-\frac{\tau }{r}+{\beta }r+(B+C\delta
_{0})\delta _{S_{12},1}~~~~~~~~~  \nonumber \\
+(-1)^{l+1}2Ae^{-\alpha r}[\vec{s_{12}}\cdot \vec{s_{3}}+\vec{t_{12}}%
\cdot \vec{t_{3}}+2\vec{s_{12}}\cdot \vec{s_{3}}~\vec{t_{12}}\cdot \vec{t_{3}%
}],
\end{eqnarray}
\noindent where $\delta_{0}$ stands in short for $\delta_{n,1}\delta_{l,0}$. This Hamiltonian is characterized by the parameters $\tau $, ${\beta }$, $E_0$,
 $A$, $\alpha$, $B$, $C$, which are chosen by comparing with experimental data.
Since all contributions are given in explicit analytic form, the
determination of the parameters is straigthforward. The procedure that we use
is the following:

\noindent
(i) the parameters ${\tau ^{2}}m$, $\frac{\beta }{m\tau }$ and $E_{0}$ 
are determined from the location of the lowest state for each orbital angular momentum; we
obtain $\tau ^{2}m=1546~\mbox{MeV}$, $\frac{\beta }{m\tau }=5~\mbox{MeV}$ and $E_{0}=1706~\mbox{MeV}$;

\noindent
(ii) the parameters $B$ and $C$ are determined by the splitting between the
state $s_{12}=0$, $t_{12}=0$ and the average of the states $s_{12}=1$, 
$t_{12}=1$. We find $B=300~\mbox{MeV}$ and $C=400~\mbox{MeV}$. The value $B~=~300~\mbox{MeV}$ is
consistent with earlier estimates, $B=250~\mbox{MeV}$, arising from an evaluation
of the color magnetic interaction \cite{deswart};

\noindent
(iii) the parameters $A$ and $\alpha$ are determined from the splitting of
the multiplet within $s_{12}=1$, $t_{12}=1$. We find $A=205~\mbox{MeV}$, $\frac{%
\alpha}{{\tau}m}=0.30$.

\begin{figure}[!ht]
{\includegraphics[width=9cm]{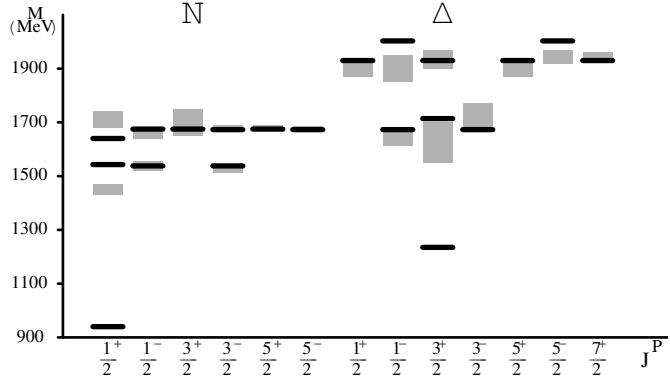}}
\caption{\small
Comparison between the calculated masses (black lines) of non-strange baryon resonances up to 2 $\mbox{GeV}$  and the experimental masses (grey boxes) from PDG \cite{pdg} 
for the 4* and 3* resonances.}
\label{spettro}
\end{figure}

\begin{table}[t]
\caption{Comparison between the calculated masses of non-strange baryon
resonances up to 2 $\mbox{GeV}$, $M_{calc}$, and the experimental masses \cite{pdg}, $%
M_{exp},$ for the 4* and 3* resonances. The masses are given in $\mbox{MeV}$. }
\label{tab:mass}
\par
\begin{center}
\begin{tabular}{ccccccccc}
\hline\hline
Baryon & Status & $M_{exp}$ && $J^{P}$&  $~l^{P}$& $~S$ & & $M_{calc}$  
\\ 
&  & (MeV) & & & &    & & (MeV) \\ 
\hline
N(939)~$P_{11}$ & **** & 938 & & $~\frac{1}{2}^+$ &  $~0^+$ & $~\frac{1}{2}$ & &  $940$ \\ 
N(1440)~$P_{11}$ & **** & 1430-1470 & & $\frac{1}{2}^+$ &  $~0^+$ & $~\frac{1}{2%
}$ &  & $1543$ \\ 
N(1520)~$D_{13}$ & **** & 1515-1530 & & $\frac{3}{2}^-$ &  $~1^-$ &$~\frac{1}{2%
}$ & &  $1538$ \\ 
N(1535)~$S_{11}$ & **** & 1520-1555 & & $\frac{1}{2}^-$& $~1^-$& $~\frac{1}{2%
}$ & &  $1538$ \\ 
N(1650)~$S_{11}$ & **** & 1640-1680 & & $\frac{1}{2}^-$ & $~1^-$ & $~\frac{3}{2%
}$ & & $1675$ \\ 
N(1675)~$D_{15}$ & **** & 1670-1685 & & $\frac{5}{2}^-$ &  $~1^-$ & $~\frac{3}{2%
}$ & &  $1673$ \\ 
N(1680)~$F_{15}$ & **** & 1675-1690 & & $\frac{5}{2}^+$ &  $~2^+$ & $~\frac{1}{2%
}$ & &   $1675$ \\
N(1700)~$D_{13}$ & *** & 1650-1750 & & $\frac{3}{2}^-$ &  $~1^-$ &  $~\frac{3}{2}
$ & &  $1673$ \\ 
N(1710)~$P_{11}$ & *** & 1680-1740 & & $\frac{1}{2}^+$ &  $~0^+$ & $~\frac{1}{2}
$ & &  $1640$ \\ 
N(1720)~$P_{13}$ & **** & 1650-1750 & & $\frac{3}{2}^+$ &  $~2^+$ & $~\frac{1}{2%
}$ & & $1675$ \\ 
\hline
$\Delta (1232)~P_{33}$ & **** & 1230-1234 & & $\frac{3}{2}^+$ & $~0^+$  & $%
~\frac{3}{2}$  & & $1235$ \\ 
$\Delta (1600)~P_{33}$ & *** & 1550-1700 & & $\frac{3}{2}^+$ & $~0^+$ & $%
~\frac{1}{2}$ & & $1714$ \\ 
$\Delta (1620)~S_{31}$ & **** & 1615-1675 & & $\frac{1}{2}^-$ & $1^-$ &  $~\frac{1}{2}$ & & $1673$ 
\\ 
$\Delta (1700)~D_{33}$ & **** & 1670-1770 & & $\frac{3}{2}^-$ & $~1^-$ & $%
~\frac{1}{2}$ & &  $1673$ \\ 
$\Delta (1900)~S_{31}$ & *** & 1850-1950 & & $\frac{1}{2}^-$ & $~1^-$ & $%
~\frac{1}{2}$ & & $2003$\\
$\Delta (1905)~F_{35}$ & **** & 1870-1920 & & $\frac{5}{2}^+$ & $~2^+$ & $%
~\frac{3}{2}$ & & $1930$ \\ 
$\Delta (1910)~P_{31}$ & **** & 1870-1920 & & $\frac{1}{2}^+$ & $~2^+$ & $%
~\frac{3}{2}$ & &  $1930$ \\ 
$\Delta (1920)~P_{33}$ & *** & 1900-1970 & & $\frac{3}{2}^+$ & $~2^+$ & $%
~\frac{3}{2}$ & &  $1930$ \\ 
$\Delta (1930)~D_{35}$ & *** & 1920-1970 & & $\frac{5}{2}^-$ & $~1^-$ & $~\frac{3}{2}$ 
& & $2003$\\
$\Delta (1950)~F_{37}$ & **** & 1940-1960 & & $\frac{7}{2}^+$ & $~2^+$  & $%
~\frac{3}{2}$ & & $1930$ \\  
\hline
\hline
\end{tabular}
\end{center}
\end{table}

In Fig. 2 and in Table II the results of the model are compared with the experimental data 
\cite{pdg}. It can be seen that the quark-diquark
model with the specific interaction, Eq.(16), provides a good description of
the masses of $4^{\ast }$ and $3^{\ast }$ resonances. The quality of this
description is similar to that of the usual three-quark model in its various
forms \cite{cap},\cite{bil}, and it can be observed that the predicted
value for the mass of the Roper resonance is higher
than the experimental data. However, in the quark-diquark model, 
some
degrees of freedom are frozen. There are therefore far fewer missing
resonances, and in particular no missing resonances in the lower part of the
spectrum. On the contrary, the problem of missing resonances plagues all
models with three constituent quarks.

It will also be noted that the clustering of states expected by the quark-diquark
model is particularly evident for $l^{P}=1^{-}$ and $l^{P}=2^{+}$, at both 
the qualitative and quantitative levels. This clustering appears both
in the nucleon, $N$, and in the $\Delta $. The clustering of states is
another feature that is difficult to obtain in a three quark description.
In the quark-diquark model, however, it is obtained automatically.


For the model described here, in which all
interactions in addition to the Coulomb-like interaction are treated in
perturbation theory, elastic and transition form factors can all be
calculated analytically. The scalar matrix elements are of the type

\begin{equation}
\int d^{3}r R_{n^{\prime },l^{\prime }}(r)Y_{l^{\prime },m^{\prime
}}(\Omega )e^{-i\vec{k}\cdot \vec{r}}R_{n,l}(r)Y_{l,m}(\Omega )~.
\end{equation}
These integrals, denoted by $U_{n~l ,n^{\prime }l ^{\prime
}}(k)\delta _{mm^{\prime }\text{ }}$ are straightforward and 
Table III shows the corresponding results for transitions from the ground state with
quantum numbers, $n=1$, $l^{P}=0^{+}$ to a state with $n, l $. It is observed that the elastic form factor is 
\begin{equation}
F(k)~=~\frac{1}{(1+k^{2}a^{2})^{2}}~~~,
\end{equation}
with $a=\frac{1}{2\tau m}$. In addition to having a power-law behavior with
momentum transfer, $k$, typical of Coulomb-like interactions, this form
factor has precisely the power dependence observed experimentally. Thus the
quark-diquark model presented here has the further advantage of producing in first approximation an
elastic form factor in agreement with experimental data, Fig. 3. All form factors
depend on the scale ${a}$. To determine the scale ${a}$ the
r.m.s. radius can be calculated by using the ground state wave functions, 
$<{r}^{2}>=\frac{3}{\tau ^{2}{m}^{2}}=12{a}^{2}$.
This calculated value is then fitted to the experimental value $%
<{r}^{2}>_{exp}=0.74(1)~\mbox{fm}^{2}$ \cite{pdg}. The resulting value is ${a}=0.25~\mbox{fm}$. Since the
parameter $\tau $ is determined from the spectrum, one obtains the reduced
mass $m=102~\mbox{MeV}$.
This value is somewhat lower than the naive expectation $%
m=200~\mbox{MeV}$, obtained by assuming the quark mass to be $300~\mbox{MeV}$ and the
diquark mass to be $600~\mbox{MeV}$. The results shown in Table III should be compared with the
analogous results in the three quark model, as for example reported in Table
IX of Ref. \cite{bil}

\begin{figure}[!ht]

\includegraphics[width=9cm]{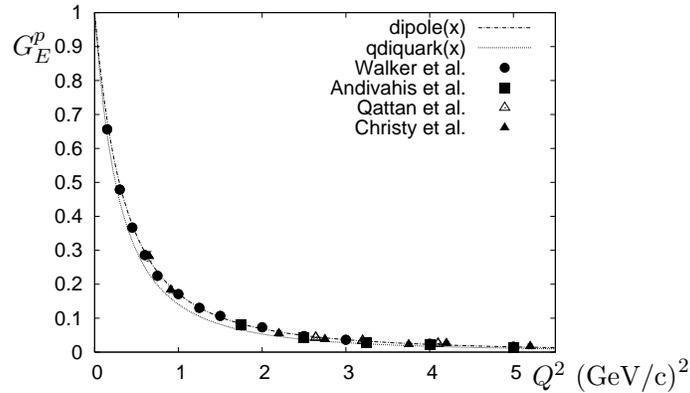}

\caption{\small
The electric form factor of the proton. The dotted line is the 
result of the model (Eq.(18)) using ${a}=0.25~\mbox{fm}$, the dot-dashed line corresponds to the dipole fit, ${a}=0.23~\mbox{fm}$. The experimental data are 
from Ref. \cite{lastdata}.
}

\label{fig:rap_comp}
\end{figure}





\begin{table}[t]
\caption{The scalar form factors of Eq.($17$) for transitions to final
states labelled by the quantum numbers $n$, $l^P$, where $P$ is the parity.
The initial state is $n=1,l^{P}=0^{+}$ and $a=\frac{1}{2\protect\tau m%
}$.}
\label{tab:ff}
\vspace{-0.4cm}
\par
\begin{center}
\begin{tabular}{cccc}
\hline
\hline
$n$ & $l^{P}$ &  & $\langle~n~l^{P} ~|~U~|~1~0^{+}~\rangle$ \\ 
\hline
$1$ & $0^+$ &  & $\frac{1}{(1+k^2a^2)^{2}}$ \\ 
$2$ & $1^-$ &  & $\frac{i}{\sqrt{2}}~(\frac{4}{9})^3~ ~\frac{24~ka} {(1+%
\frac{16}{9}k^2a^2)^{3}}$ \\ 
$2$ & $0^+$ &  & $16\sqrt{2}(\frac{4}{9})^3~\frac{(ka) ^2} {(1+\frac{16}{9}k^2a^2)^{3}}
$ \\ 
$3$ & $2^+$ &  & $-\frac{4}{\sqrt{6}}~(\frac{9}{16})^2~ \frac{(ka)^2} {(1+%
\frac{9}{4}k^2a^2)^{4}}$ \\ 
$3$ & $1^-$ &  & $~~\frac{i\sqrt{2}~64~ka}{27}~(\frac{9}{16})^3~ \frac{(1+~%
\frac{27}{4}~(ka)^2)} {(1+\frac{9}{4}k^2a^2)^{4}}$ \\ 
$3$ & $0^+$ &  & $\frac{4}{\sqrt{3}}~(\frac{9}{16})^2~ \frac{(1+\frac{27}{4}%
~(ka)^2)(ka)^2} {(1+\frac{9}{4}k^2a^2)^{4}}$ \\ 
\hline \hline
\end{tabular}
\end{center}
\end{table}

In order to calculate the magnetic elastic form factors and the helicity 
amplitudes, other matrix elements are also needed to be calculated; this program will be completed in another article (not least because a relativistic version of the model is required for a good calculation of form factors), since here 
we are interested only to explore the qualitative features of the model and of its results.
In particular the quark-diquark model presented here produces the phenomenon of
stretching, which is at the basis of the Regge behavior of hadrons. The
transition radii increase with $n$ and $l$, as one can see from Table III, or
by evaluating 
 $~<r^{2}>_{n,l=n-1}=(2n+2)(2n+1)\frac{n^{2}}{4{\tau }^{2}{m}^{2}}$.
In other words, hadrons swell as the angular momentum increases.



In this article, we present a simple quark-diquark model with a
specific direct plus exchange interaction. This model reproduces the
spectrum just as well as conventional three-quark models. However, 
it has far fewer missing resonances than the usual models. Most importantly, the model
produces form factors with power-law behavior as a function of momentum
transfer, in agreement with experimental data. Finally, it shows the phenomenon of
stretching which is at the basis of the Regge behaviour of hadrons.


One may wonder whether there is a unique spectral signature for
quark-diquark models. One of these signatures is the detection of ${1^{+}}$
states which are antisymmetric in all three quarks. These states,
originating from the omitted diquark representation $\bf{15}$ of $SU_{sf}(6)$ are not present in
the quark-diquark model and occur (at different masses) in all models with
three quarks. These missing states may, however, be very difficult to detect
since they are decoupled and cannot be excited with electrons or photons. To
excite these states, strongly interacting particles are needed, for example $(%
\vec{p},\vec{p^{\prime }})$ with spin transfer. Another possibility is to
study whether or not the mixed symmetry states (which are doubly degenerate
in the $q^{3}$ model) are in fact simply degenerate (as in the quark-diquark
model). 

The work presented here can be expanded in several directions: (1) to
include strange baryons; this expansion is straightforward and requires no
further assumption; (2) to study multiquark states;
 (3) to include relativistic corrections.

For transparency, here we have presented results in which all interactions,
in addition to the Coulomb-like force, are treated in perturbation theory.
Numerical diagonalization shows that with the parameters of this article,
the perturbation treatment is valid (at least for the low-lying states).
The complete numerical results will be the subject of a subsequent
paper.

An aspect that has not been discussed here is how to derive the
quark-diquark model from microscopy. This aspect has been the subject of
many investigations in the similar problems of condensed matter and nuclear
physics. The usual argument in hadronic physics is that diquark correlations
 arise from the spin-spin interaction originating from one gluon exchange 
that lowers the scalar diquark relative to the vector diquark.
 Some interesting results are now
available and show that instanton interactions favor diquark clustering
 \cite{Shuryak}, and that by using an Instanton Liquid Model these generate a deeply bound
scalar antitriplet diquark not point-like \cite{Faccioli}.
In this respect instantons can provide a microscopic dynamical mechanism in terms of non perturbative QCD interactions. 
In this article, we have merely tried to describe many data by means 
 of a phenomenological approach; the objective of a subsequent paper will be to understand what kind of microscopic QCD mechanisms we are trying to mimic in this oversimplified form.



\end{document}